\documentclass[aps,pra,twocolumn,eqsecnum,amsmath,superscriptaddress,showpacs]{revtex4}

\usepackage{graphicx}
\usepackage{longtable}
\usepackage{epsfig}
\usepackage{dcolumn}
\usepackage{bm}
\usepackage{amssymb}
\usepackage{multirow}
\usepackage{times,color}
\usepackage{hyperref}

\begin{document}

\title{Generalized fidelity susceptibility at phase transitions}

\author {Wen-Long You }
\affiliation{College of Physics, Optoelectronics and Energy, Soochow
University, Suzhou, Jiangsu 215006, People's Republic of China}

\author {Li He}
\affiliation{Jiangsu University of Science and Technology, Zhangjiagang, Jiangsu 215600, China}

\date{\today}

\begin{abstract}
In the present work, we investigate the intrinsic relation between quantum fidelity susceptibility (QFS) and the dynamical structure factor. We give a concise proof of the QFS beyond the perturbation theory. With the QFS in the Lehmann representation, we point out that the QFS actually the negative-two-power moment of dynamical structure factor, and illuminate the inherent relation between physical quantities in the linear response theory. Moreover, we discuss the generalized fidelity susceptibility (GFS) of any quantum relevant operator, that may not be coupled to the driving parameter, present similar scaling behaviors. Finally, we demonstrate that the QFS can not capture the fourth-order quantum phase transition in a spin-1/2 anisotropic XY chain in the transverse alternating field, while a lower-order GFS can seize the criticalities.
\end{abstract}

\pacs{05.70.Jk,64.70.Tg ,71.10.Fd,71.45.Gm}

\maketitle
 \section{Introduction}
\label{sec:introduction}
Recently the concept of quantum fidelity susceptibility (QFS) has been
recognized as a versatile indicator in identifying quantum critical points (QCPs) \cite{Gu10}. Plentiful literatures reveal that the QFS exhibits singularity at a great variety of critical points \cite{Yang08,Chen08,Yu09,Cheng10,Mukherjee11,Thakurathi12,Damskim13,Luo14}. Nevertheless, there still remains vague issues in understanding the QFS. For a many-body Hamiltonian $\hat{H}(\lambda)=H_0 + \lambda H_I$,
when the driving parameter $\lambda$ varies across a transition point $\lambda_c$, the noncommutability between $H_0$ and $H_I$ leads to a nonanalytic behavior in the ground state $\Psi_0$ and a reconstruction of low-energy spectra $E_l$. Originated from quantum information theory,
quantum fidelity is defined as the overlap of wavefunctions at two points in parameter space:
 \begin{eqnarray}
F(\lambda_0,\lambda_1)=\sqrt{  \langle \Psi_0 (\lambda_0) \vert \Psi_0   (\lambda_1) \rangle\langle \Psi_0 (\lambda_1) \vert \Psi_0   (\lambda_0) \rangle }.
\end{eqnarray}
It was discovered that quantum criticalities
promote the decay of fidelity \cite{Quan06}. Such sensitivity to the quantum criticality was further amplified by the associated changing rate.
In this respect, the QFS is defined as the leading-order of the Taylor expansion of the overlap function $F(\lambda,\lambda+\delta \lambda)$ \cite{You07},
 \begin{eqnarray}
\chi_{\rm F}= \lim_{\delta \lambda \to 0} \frac{-2 \ln F(\lambda,\lambda+\delta \lambda)}{(\delta \lambda)^2}. \label{chiform}
\end{eqnarray}
As $\lambda$ crosses QCPs adiabatically, $\chi_{\rm F}$ shows a peak singularity and exhibit universal scaling law. Such universal properties of QCPs in equilibrium prompted
a rapidly growing theoretical interest to the quench dynamics.

When the tuning parameter $\lambda$ is time dependent, say,
$\lambda(t) \sim v t^r $, where
$v$ is the adiabatic parameter that controls the proximity to the instantaneous ground state, 
 the energy gap $\Delta$ at the critical point $\lambda_c$ tends to zero when the size of the system $N$ $\to$ $\infty$ and it is almost impossible to pass the critical point at a finite speed $v$ without exciting the system.
The nonanalytic scalings can be revealed by singularities in the generalized adiabatic susceptibility $\chi_{2r+2}(\lambda)$ of order $2r+2$ \cite{Grandi}:
 \begin{eqnarray}
 \chi_{2r+2}= \sum_{l\neq 0} \frac{\vert \langle  \Psi_l (\lambda) \vert H_I \vert \Psi_0 (\lambda)\rangle \vert^2}{[E_l(\lambda)-E_0(\lambda)]^{2r+2}},
\label{generalizedsusceptibility}
\end{eqnarray}
where $\Psi_l$ and $E_l$ ($l \neq 0$) correspond to the excited states and energies respectively.
The general theory for crossing the critical point at a finite rate is given by Kibble-Zurek mechanism \cite{Kibble,Zurek}.
In the opposite limit, i.e., $v$$ \to$ 0,
Ref. [\onlinecite{You07}] firstly discussed the adiabatic evolution of ground state in the frame of perturbation theory, and obtained
\begin{eqnarray}
\chi_{\rm F}= \sum_{l\neq 0} \frac{\vert \langle  \Psi_l (\lambda) \vert H_I \vert \Psi_0 (\lambda)\rangle \vert^2}{[E_l(\lambda)-E_0(\lambda)]^2}. \label{chiperturbationform}
\end{eqnarray}
This exactly corresponds to the generalized adiabatic susceptibility [Eq.(\ref{generalizedsusceptibility})] of order $2$.
Note that the second derivative of the ground-state energy is of order $1$.

Due to the arbitrariness of relevance of the driving Hamiltonian $H_I$ under the renormalization group transformation, $\chi_F$ increases as the system size grows and manifests distinct scaling behavior.
The summation in Eq. (\ref{chiperturbationform}) contributes to an extensive scaling of $\chi_{\rm F}$ in the off-critical region. Therefore, QFS per site
$\chi_{\rm F}/N$ appears to be a well-defined value, where $N$ $\equiv$ $ L^d$ is
the number of sites and $d$ stands for system dimensionality.
Instead, the QFS exhibits much stronger dependence on system size across
the critical point than in noncritical region, showing that a
singularity emerges in the summation of Eq. (\ref{chiperturbationform}). This implies an abrupt change in the ground state of the system at the QCP in the thermodynamic limit. For example, when $H_I$ is charge-or spin current operator $\hat{J}$ or the polarization operator $\hat{P}$, it is shown that in gapless one-dimensional (1D) systems with open boundary conditions  the leading terms in the $L$ dependence are given by
$\chi_{\hat{J}}$ $\propto$ $KL^2$
and
$\chi_{\hat{P}}$  $\propto$
$KL^4/u^2$, where
$K$
is the Luttinger liquid parameter,
$u$ is the
characteristic ¡°sound¡± velocity, and the numerical prefactors
are universal \cite{Greschner13}. Following standard arguments in scaling analysis of a second-order quantum phase transition (QPT), one obtains that the QFS per site in $d$ dimensions in most cases ($\nu d>2$) scales as \cite{Schwandt09,Grandi,Marek}
\begin{eqnarray}
 &&\chi_F /N \sim L^{2/\nu-d} f(\vert \lambda-\lambda_c \vert L^{1/\nu}), \label{2ndQPTchiscaling}
\end{eqnarray}
where $\nu$ is the critical exponent of the correlation length and $f(\cdot)$ is a
regular scaling function. As for the first-order QPT, the QFS per site scales
exponentially \cite{You11}, i.e.,
 \begin{eqnarray}
 \chi_F /N \sim g(N)e^{\mu N}, \label{1stQPTchiscaling}
 \end{eqnarray}
 where $\mu$ is a size-independent constant, and $g(N)$ is a
polynomial function of $N$.

In addition to an increasing interest in the finite-size scaling analysis, a few researches addressing that the QFS cannot detect the QPTs of infinite order, have joined in
the subject of active study \cite{Venuti07,Chen08,Yu14}. This conclusion conflicts with the findings on the 1D Luttinger model \cite{Yang07,Fjarestad08}, and the XXZ chain by virtue of the density-matrix renormalization-group technique \cite{Bo10} and the real-space quantum renormalization group \cite{Langari12}, but supports initial opinion proposed by You {\it et al.} \cite{You07}.
However, recently G. Sun {\it et al.} argued that using
the QFS as a sensor to detect Berezinskii-Kosterlitz-Thouless-type transitions is uncontrollable \cite{Sun14}. Therefore, the issue on fidelity susceptibility in describing high-order continuous phase transitions seems still controversial.

Contrary to the QFS, the thermal-state fidelity is determined as a descendant of Uhlmann fidelity \cite{Uhlmann}
 \begin{eqnarray}
F_T(\rho_0,\rho_1)={\rm tr} \sqrt{\sqrt{\rho_1}\rho_0\sqrt{\rho_1}},
\end{eqnarray}
where $\rho_i$ ($i$=$0, 1$) are density
operators. A simple formula is achieved by considering temperature driven phase transitions \cite{Zanardi07b},
\begin{eqnarray}
F_T(\beta_0,\beta_1)=\frac{Z[(\beta_0+\beta_1)/2]}{\sqrt{Z(\beta_0)Z(\beta_1)}}. \label{F-Thermal}
\end{eqnarray}
Then there is a direct connection between the thermal fidelity susceptibility and a purely thermodynamic quantity \cite{You07}, namely, specific heat,
\begin{eqnarray}
\chi_{T}=\frac{C_v}{4 \beta^2}. \label{F-Thermal2}
\end{eqnarray}
Eq. (\ref{F-Thermal2}) can not fall to the QFS in zero-temperature limit, $\beta \to \infty$. Meanwhile, the QFS for degenerate ground state is ill defined.

In the present work, we analyse the immanent relation between the QFS and the dynamical structure factor. We give a concise proof of Eq.(\ref{chiperturbationform}) beyond the perturbation theory in Sec.\ref{sec:nonperturbative}. With the QFS in such Lehmann representation, we point out that the QFS actually the negative-two-power moment of dynamical structure factor in Sec.\ref{sec:Dynamic}, and illuminate the inherent relation between physical quantities in the linear response theory. In Sec.\ref{sec:application},
we discuss the critical behavior of the generalized fidelity susceptibility (GFS) in 1D transverse Ising model. We also show that the QFS fails in detecting the QPT in 1D anisotropic XY model under a transverse alternating field, while a lower order GFS ${\cal M}_{(-3)}$ can probe the fourth-order QPT therein. The paper is summarized in Sec. \ref{sec:conclusion}.

\section{Fidelity susceptibility in nonperturbative form}
 \label{sec:nonperturbative}
The universal infrared divergence comes from the low-frequency contribution in Eq. (\ref{chiperturbationform}).
This prompts us to consider the dynamical response of the system to the interaction $\partial_\lambda H$'s within the adiabatic perturbation theory. To this end, the susceptibilities can be also expressed through the imaginary time correlation function of the perturbation $H_I$ ($\tau$) using the relation \cite{You07,Venuti07}:
\begin{eqnarray}
\chi_F =\int_{0}^{\infty}\tau\left[ \langle \Psi_0 |H_I(\tau) H_I(0)|\Psi_0\rangle
-\langle\Psi_0|H_I|\Psi_0\rangle^2\right] d\tau,\nonumber \\
\label{eq:fidelityimag}
\end{eqnarray}
where $H_I(\tau)$=$\exp(H \tau) H_I \exp(-H \tau)$.
In this spirit, the QFS was extended straightforwardly to finite temperature \cite{Schwandt09,Albuquerque10}:
\begin{eqnarray}
\chi_F(\beta)=\int_{0}^{\frac{\beta}{2}} \tau\left[\langle \Psi_0 |H_I(\tau) H_I |\Psi_0\rangle -\langle\Psi_0|H_I|\Psi_0\rangle^2\right] d\tau,\nonumber \\
\label{eq:fidelityfnal}
\end{eqnarray}
in which $\beta$ is the inverse temperature.

In the Lehmann representation [Eq.(\ref{chiperturbationform})], the hopping matrix between ground state and arbitrary excitations is written as
\begin{eqnarray}
\langle \Psi_l \vert H_I \vert \Psi_0 \rangle= \langle \Psi_l \vert \partial_\lambda H \vert \Psi_0 \rangle  = (E_0-E_l) \langle \Psi_l \vert\partial_\lambda \Psi_0 \rangle.
\end{eqnarray}
Putting the above relation into right hand of Eq.(\ref{chiperturbationform}) and making use of completeness relation $\sum_l \vert \Psi_l \rangle \langle \Psi_l \vert=1 $, we obtain
\begin{eqnarray}
 \sum_{l\neq 0} \frac{\vert \langle  \Psi_l \vert H_I \vert \Psi_0 \rangle \vert^2}{(E_l-E_0)^2} = \langle \partial_\lambda \Psi_0 \vert \partial_\lambda \Psi_0 \rangle - \vert  \langle \partial_\lambda \Psi_0\vert \Psi_0 \rangle \vert^2. \label{chipartialform2}
\end{eqnarray}
 Remarkably, the QFS was now devised to seize the criticality in the perspective of the Riemannian metric tensor form\cite{Zanardi07a}
\begin{eqnarray}
\chi_{\rm F}=\langle \partial_\lambda \Psi_0 \vert \partial_\lambda \Psi_0 \rangle - \vert  \langle \partial_\lambda \Psi_0 \vert \Psi_0 \rangle \vert^2. \label{chipartialform}
\end{eqnarray}
The implications of the equivalence between Eq.(\ref{chipartialform}) and Eq.(\ref{chiperturbationform}) are three fold. First, it shows that the QFS is well defined as a consequence of orthogonality of energy levels, without prior knowledge of fidelity. Second, the usage of Eq.(\ref{chipartialform}) is nonperturbative, so the adiabatical perturbation theory is not indispensable. Last but not the least, the description of Riemannian manifold abandons the requirement that the parameter $\lambda$ is coupled with the driving Hamiltonian. In other words, we can investigate the fidelity susceptibility of any quantum operator $\hat{O}$, that may not appear in the Hamiltonian explicitly. Such tentative investigation has been demonstrated in Ref.[\onlinecite{Thesberg11}].

\section{Dynamic structure factor}
\label{sec:Dynamic}
A most common experimental characterization in condensed matter physics is inelastic neutron scattering, which directly measures the energy and
momentum dependence of spin-spin correlations as described by the momentum-resolved dynamical structure factor. For a quantum operator $\hat{O}$, the momentum
integrated dynamic structure factor $S(\omega)$ is defined as\cite{Enrico}
\begin{eqnarray}
S(\hat{O}, \omega)= \sum_{l \neq 0}\vert \langle \Psi_l \vert \hat{O} \vert \Psi_0 \rangle \vert^2 \delta(\omega-\omega_{l0}).
\end{eqnarray}
Here $\omega_{l0}= E_l-E_0$ denotes the corresponding excitation
energies of $l_{\rm th}$ excited state. $S(\omega)$ contains rich information of a many-body system. For instance, the dynamic response function in linear response theory can be expressed as\cite{Enrico}
\begin{eqnarray}
\chi(\hat{O}, \omega)=\int_0^{\infty} d \omega' \left[ \frac{S(\hat{O}, \omega')}{\omega+\omega'+i 0 ^+ }-\frac{S(\hat{O}^\dagger,\omega')}{\omega-\omega'+i 0 ^+ } \right], \label{DRF}
\end{eqnarray}
which can be also linked with experimental signal by applying external perturbing field through Kubo formula. If $\hat{O}$ is Hermitian or the system is time-reversal invariant, as is supposed in the following, $\vert \langle \Psi_l \vert \hat{O} \vert  \Psi_0 \rangle \vert^2$ $=$ $\vert \langle \Psi_l \vert \hat{O}^{\dagger} \vert  \Psi_0 \rangle \vert^2$, then Eq.(\ref{DRF}) takes on a simplified form:
\begin{eqnarray}
\chi(\hat{O}, \omega)=\int_0^{\infty} 2 \omega' d \omega'  \left[ \frac{S(\hat{O}, \omega')}{\omega'^2-(\omega+i 0 ^+)^2 } \right].
\end{eqnarray}

The linear response function can be related to the moments ${\cal M}_{(k)}$ of the dynamic form factor, defined by
\begin{eqnarray}
{\cal M}_{(k)}(\hat{O})=\int_0^{\infty} d \omega \omega^k S(\hat{O},\omega)=\sum_{l} \omega^k_{l0} \left\vert \langle \Psi_l \vert \hat{O} \vert \Psi_0 \rangle \right\vert^2,   \label{Mkdefinition}
\end{eqnarray}
where $k$ is an integer. The moments of different orders give rise to various sum rules and have explicit physical meanings. For example, zero-order moment, also called as static structure factor, quantifies the fluctuation at zero temperature,
\begin{eqnarray}
{\cal M}_{(0)}(\hat{O})=\langle \Psi_0 \vert {\hat{O}}^2\vert \Psi_0 \rangle - \vert \langle \Psi_0 \vert \hat{O} \vert \Psi_0 \rangle \vert^2, \label{Mktemperature}
\end{eqnarray}
which is the leading response in linear quench dynamics \cite{YU12}. Similarly, negative-one-order moment corresponds to static response function
\begin{eqnarray}
{\cal M}_{(-1)}(\hat{O})= \frac{1}{2}\chi(\hat{O},\omega=0).
\end{eqnarray}
From Eq.({\ref{chipartialform}), negative-two-order moment is the QFS with the relation
\begin{eqnarray}
 {\cal M}_{(-2)}(\hat{O})=\chi_F(\hat{O}). \label{chimomentumform}
\end{eqnarray}
Eq.(\ref{chimomentumform}) bridges the QFS and dynamic structure factor, and enriches the techniques to obtain the QFS. In other words, Eq. (\ref{Mkdefinition}) can be seen as a generalization of the QFS, and
${\cal M}_{(k)}$ is named as $k$-order GFS. Although $S(\hat{O}, \omega)$ is not easy to retrieve, the dynamic structure factor $\chi(\omega)$ can be solved in terms of a series of Feynman diagrams, and thus the dynamical structure factor can be gotten from Eq.(\ref{DRF}):
 \begin{eqnarray}
S(\hat{O},\omega)=\frac{{\rm Im} \chi(\hat{O}, \omega)}{\pi}, \quad (\omega>0).
\end{eqnarray}
As far as $S(\hat{O}, \omega)$ is known, finite-order GFSs can be gained.

For a quantum operator $\hat{O}$ at
equilibrium at a finite temperature $T$, the momentum
integrated dynamic structure factor $S_T(\omega)$ is defined as
\begin{eqnarray}
S_T(\hat{O}, \omega)= \sum_{l \neq n} \frac{e^{-\beta E_n}}{Z} \vert \langle \Psi_l \vert \hat{O} \vert \Psi_n \rangle \vert^2 \delta(\omega-\omega_{ln}), \label{Mktemperature2}
\end{eqnarray}
where the summation runs over both $l$ and $n$ excluding they are equal. Here partition function $Z =\sum_l e^{-\beta E_n}$ and $\omega_{ln}= E_l-E_n$ denotes the
energy difference between excited states $\vert l \rangle$ and $\vert n \rangle$. In the zero-temperature limit, $\vert n \rangle$ is restricted to ground state $\vert 0 \rangle$, and the dynamic
form factor coincides for positive $\omega$, since only processes by which energy is transferred to the system are allowed. At finite temperature, the corresponding GFS of the dynamic form factor
\begin{eqnarray}
{\cal M}_{(k)}((\hat{O}))&=&\int_{-\infty}^{\infty} d \omega \omega^k S_T(\hat{O},\omega)\nonumber \\
&=&\sum_{n\neq l} \omega^k_{ln} \left\vert \langle \Psi_l \vert \hat{O} \vert \Psi_n \rangle \right\vert^2. \label{Mktemperature3}
\end{eqnarray}
Note that these GFSs, contrary to the $T = 0$ case, get contribution from negative $\omega$ as well, since transferring energy both from the particle to the system, or from the system to the rest particle are both admissible at $T > 0$.

Much attention has been paid to the critical behavior of correlation function in the QPT. In the low-frequency limit, the dominated contribution comes from static response function $\chi$. Provided $\chi$ shows a clear singularity, such anomaly can act as a signature of criticality. While for some phase transitions, $\chi$ does not manifest singular trend, and then ${\cal M}_{(-2)}$, equivalently the QFS, can be used as a supplementary means of identifying the phase transitions. The QFS was proven to successfully characterize the critical behavior of QPTs in a many-body system \cite{Gu10}.

Common speaking, the reconstruction of low-energy excitations mainly contributes to the negative-order GFS while the change in high-energy part is crucial to positive-order GFS. When quantum operator $\hat{O}$ acting on ground state has nonzero components with low-energy states, the singularity is more pronounced for larger magnitude of $\vert k \vert$. In fact, for arbitrary $k$, we have
\begin{eqnarray}
\omega_{\rm min} \le \frac{{\cal M}_{(k+1)}}{{\cal M}_{(k)}},
\end{eqnarray}
where $\omega_{\rm min}$ is the lowest excitation. The equality only holds for the existence of $\Psi_{\rm min}$ and $\langle \Psi_{\rm min} \vert \hat{O} \vert \Psi_0 \rangle =1$. It is clear that the lower power of negative GFS displays more divergence provided gap closes at criticality.
\section{Application and results}
\label{sec:application}
To see such generalization in Eq. (\ref{Mkdefinition}) plays an essential roles in various types of QPTs, we calculate the GFSs in a few examples in the following. There has been intense
interest in studying 1D spin systems such as transverse Ising model, which serve as a general prototype for quantum magnetism and for the understanding of QPTs. All the concepts about both equilibrium and non-equilibrium QPTs have been tested on this
model. The 1D spin-1/2 transverse Ising model is well known for its solvability and paradigm, given by \cite{Suzuki}
\begin{eqnarray}
\hat{H}=-\sum_{j=1}^{N} (\sigma_{j}^{x}\sigma_{j+1}^{x}+h \sigma_{j}^{z}), \label{1DIsingModel}
\end{eqnarray}
where $h$ is the strength of transverse field and $N$ is the number of spins (assumed here to be even). The Hamiltonian (\ref{1DIsingModel}) respects the Kramers-Wannier duality symmetry, reflected in that the site in the original chain has one-to-one correspondence with the bond of the dual chain and vice versa by a non-local mapping:
\begin{eqnarray}
\tau_{j}^{x}&=&\sigma_{j}^{z} \sigma_{j+1}^{z},
 \label{taujx}    \\
\tau_{j}^{z}&=& \prod_{k \le j} \sigma_{k}^{x}. \label{taujz}
\end{eqnarray}
The operators defined in Eqs. (\ref{taujx})-(\ref{taujz}) satisfy the same set of commutation relation as the Pauli operators, i.e., they
commute on different sites and anti-commute on the same site. The Jordan-Wigner transformation maps explicitly between pseudospin operators and spinless
fermion operators by
\begin{eqnarray}
\sigma _{j}^{+}& =&\exp\left[ i\pi\sum_{i=1}^{j-1}c_{i}^{\dagger }c_{i}^{}
\right] c_{j}^{}=\prod_{i=1}^{j-1}\sigma _{i}^{z}c_{j}^{}, \label{JW1}   \\
\sigma _{j}^{-}& =&\exp\left[-i\pi\sum_{i=1}^{j-1}c_{i}^{\dagger }c_{i}^{}
\right] c_{j}^{\dagger}=\prod_{i=1}^{j-1}\sigma_{i}^{z}c_j^{\dagger }, \\
\sigma _{j}^{z}& =&1-2c_{j}^{\dagger }c_{j}^{}. \label{JW2}
\end{eqnarray}
After the fermionization, Fourier transformation can be proceeded consequently,
\begin{eqnarray}
c_{j}&=&\frac{1}{\sqrt{N}}\sum_{q}e^{-iq j}c_{q},
\end{eqnarray}
with the discrete momenta $q$, which correspond to the reduced Brillouin zone and are given by
\begin{equation}
q=\frac{n\pi}{N}, \quad n= -(N-1), -(N-3),
\ldots,(N-1).
\label{kset}
\end{equation}
Hereupon the Bogoliubov-de Gennes (BdG) Hamiltonian is obtained as
\begin{eqnarray}
\hat{H}=-\sum_{q}\left[(2 \cos q-2 h)c_q^\dagger c_q + i \sin q (c_{-q}^\dagger c_{q}^\dagger + c_{-q} c_{q}) \right].\label{IsingFT}
\end{eqnarray}
Bogoliubov transformation can be adopted to diagonalize the quadratic Hamiltonian via
\begin{eqnarray}
c_q&=&\cos \theta_q b_q + i \sin \theta_q b_{-q}^{\dagger}, \nonumber \\
c_q^{\dagger}&=&\cos \theta_q b_q^{\dagger} - i \sin \theta_q b_{-q},\nonumber \\
c_{-q}&=&\cos \theta_q b_{-q} -i \sin \theta_q b_{q}^{\dagger}, \nonumber \\
c_{-q}^{\dagger}&=&\cos \theta_q b_{-q}^{\dagger} + i \sin \theta_q b_{q},
\end{eqnarray}
where $b_{q}$ and $b_{q}^{\dagger}$
are fermionic operator satisfying the same anticommutation
relation as $c_q$ and $c_q^{\dagger}$. The coefficients are determined
by
\begin{eqnarray}
\cos 2\theta_q&=&\frac{\cos q - h}{\omega_q}, \nonumber \\
\sin 2\theta_q&=&\frac{\sin q}{\omega_q},
\end{eqnarray}
where $\omega_q=\sqrt{1+h^2-2h \cos q}$. Such that the Hamiltonian becomes a quasi-free fermion system
\begin{eqnarray}
\hat{H}= \sum_{q} 2\omega_q ( b_q^{\dagger} b_q -\frac{1}{2}).
\end{eqnarray}
Obviously the ground state energy is obtained,
\begin{eqnarray}
E_0 = - \sum_{q} \omega_q. \label{E0expression}
\end{eqnarray}
There is an energy gap in the excitation spectrum of the system which goes to zero at $q=0$ for $h=1$, i.e.,
\begin{eqnarray}
\Delta(h)=E_1-E_0=2\vert 1-h \vert, \label{gap}
\end{eqnarray}
indicating the divergence of the correlation length and a QPT at $h_c$ = 1 from an ordered state ($\langle \sigma^x \rangle$ = 1) to a disordered state ($\langle \sigma^x \rangle$ = 0), which is the consequence of ferromagnetic-paramagnetic duality. To investigate the field induced criticality, a natural choice
is $z$-component magnetization $\hat{O}$=$\hat{\sigma}^z$. The dynamical structure factor of spin magnetization yields
\begin{eqnarray}
S(\hat{\sigma}^z,\omega)=\sum_q \frac{\sin^2 q}{\omega_q^2} \delta(\omega-2\omega_q).
\end{eqnarray}
The static response function takes on
\begin{eqnarray}
\chi(\hat{\sigma}^z,0)&=&2 \int_0^{\infty} d \omega \frac{S(\hat{\sigma}^z, \omega)}{\omega} =\sum_q \frac{\sin^2 q}{\omega_q^3}. \label{magneticsusofsigmaz}
\end{eqnarray}
Away from critical point $h \neq 1$, we have a linear scaling as
\begin{eqnarray}
\chi(\hat{\sigma}^z,0)= \frac{(1+h^2)K[\alpha]-(1+h)^2 E[\alpha]}{\pi h^2(1+h)} N,
\end{eqnarray}
where $K[\alpha]$ and $ E[\alpha] $ correspond to the first and second kind complete elliptic integrals of $\alpha \equiv  4h/(1+h)^2 $. Around the critical point $h=1$, we have
\begin{eqnarray}
&&\frac{\chi(\hat{\sigma}^z,0)}{N}  \sim  \ln \vert h-1 \vert, \nonumber \\&& \chi(\hat{\sigma}^z, 0)\vert_{h \to 1} \sim  N \ln N.
\end{eqnarray}
The QFS is simultaneously given by
\begin{eqnarray}
\chi_F(\hat{\sigma}^z)=\int_0^{\infty} d \omega \frac{S(\hat{\sigma}^z,\omega)}{\omega^2} =\sum_q \frac{\sin^2 q}{8 \omega_q^4}, \label{FSofsigmaz}
\end{eqnarray}
The result (\ref{FSofsigmaz}) is identical with that from  Eq.(\ref{chipartialform}) directly. A closed form of the QFS was revealed in Ref.[\onlinecite{Damski13}]:
 \begin{eqnarray}
\chi_{\rm F}(\hat{\sigma}^z)=\frac{N^2}{16h^2} \frac{h^N}{(h^N+1)^2}+\frac{N}{16h^2} \frac{h^N-h^2}{(h^N+1)(h^2-1)}. \nonumber \\
\end{eqnarray}
Away from critical point $h \neq 1$, we have
\begin{eqnarray}
 &&\frac{\chi_{\rm F}(\hat{\sigma}^z, 0<\vert h \vert<1 )}{N}= \frac{1}{ 16(1-h^2) },  \\&&\frac{\chi_{\rm F}(\hat{\sigma}^z,\vert h\vert > 1)}{N}= \frac{1}{ 16 h^2(h^2-1) } ,
\end{eqnarray}
and around critical point,
\begin{eqnarray}
 && \chi_{\rm F}(\hat{\sigma}^z,h \to 1) = \sum_{q=\frac{\pi}{N}}^{(\pi-\frac{\pi}{N})} \frac{\cot^2(q/2)}{16} \propto (\frac{1}{q})^2 \sim  N^2, \nonumber \\
\end{eqnarray}
where the lower bound gives the leading contribution in the cotangent series. Thus $\nu=1$ can be retrieved by Eq.({\ref{2ndQPTchiscaling}). Particularly, Kramers-Wannier duality of the Ising model grantees the relation:
\begin{eqnarray}
h^2 \chi_{\rm F}(\hat{\sigma}^z,h) = (\frac{1}{h})^2 \chi_{\rm F}(\hat{\sigma}^z,\frac{1}{h}).
\end{eqnarray}
\begin{figure}[h]
\includegraphics[width=8cm]{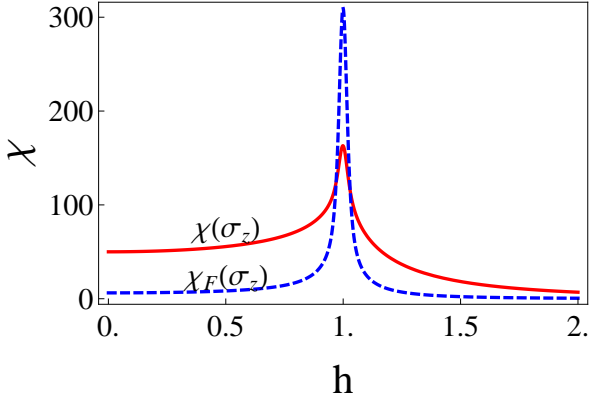}
 \caption{(Color online) The magnetic susceptibility (\ref{magneticsusofsigmaz}) and fidelity susceptibility (\ref{FSofsigmaz}) as a function of magnetic field $h$ on $N=100$ transverse Ising chain. }\label{chi-h}
\end{figure}
Figure \ref{chi-h} shows that apparently the QFS exhibits more pronounced singularity than magnetic susceptibility. Remarkably, Ref.[\onlinecite{Gu14}] disclosed a potential measurement of the QFS through the zero-momentum dynamical structure factor.

There are other choices of $\hat{O}$, e.g., momentum space Majorana operator $\hat{O}_{\rm MF}=\sum_q (c_q^{\dagger}+c_{-q})$. The dynamical structure factor of Majorana operator yields
\begin{eqnarray}
S(\hat{O}_{\rm MF},\omega)=\sum_q  \delta(\omega-\omega_q).
\end{eqnarray}
Accordingly,
\begin{eqnarray}
\chi_F(\hat{O}_{\rm MF}) &=& \sum_q \frac{\vert \cos \theta_q+ i\sin \theta_q \vert^2}{\omega_q^2}= \sum_q \frac{1}{\omega_q^2} .\
\end{eqnarray}
The asymptotic behavior appears that
\begin{eqnarray}
&&\frac{\chi_F(\hat{O}_{\rm MF})}{N} \sim  \frac{1}{\vert 1- h^2 \vert}, \quad h\neq 1, \nonumber\\
&&\chi_F(\hat{O}_{\rm MF}, h \to 1)  \sim  N^2.
\end{eqnarray}

A natural question is that whether we can find operator $\hat{O}$, whose scaling exhibits stronger singularity. It is straightforward to understand it is impossible to find an observable operator $\hat{O}$, which is not related to driving parameter $\lambda$ and exhibits more singular QFS than that of $\hat{O}_{\rm MF}$, because the denominator of Eq.(\ref{chiperturbationform}) is at most inverse square of gap (\ref{gap}), that is,
\begin{eqnarray}
\chi_{F}(\hat{O}) \le \frac{1}{\Delta^2}\sum_{l\neq 0} \vert \langle  \Psi_l (\lambda) \vert \hat{O} \vert \Psi_0 (\lambda)\rangle \vert^2.
\end{eqnarray}

\begin{figure}[h]
\includegraphics[width=8cm]{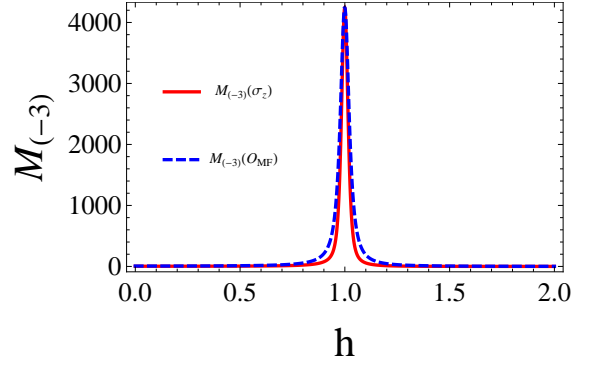}
 \caption{(Color online) The GFSs ${\cal M}_{(-3)}$ of $\hat{\sigma}^z$ and Majorana operator $\hat{O}_{\rm MF}$ as a function of magnetic field $h$ on $N=100$ transverse Ising chain. ${\cal M}_{(-3)}(\hat{O}_{\rm MF})$ has been divided by 16 to make it comparable with ${\cal M}_{(-3)}(\hat{\sigma}^z)$.}\label{Mm3-h}
\end{figure}

Provided $\chi_{\rm F}$ is not sensitive enough to observe a QPT, GFSs ${\cal M}_{(k)}$ with more negative $k$ ($k<-2$) is needed, such that
\begin{eqnarray}
{\cal M}_{(-3)}(\hat{\sigma}^z) &=&   \sum_q \frac{\sin^2 q}{16\omega_q^5}, \\
{\cal M}_{(-3)}(\hat{O}_{\rm MF}) &=&   \sum_q \frac{1}{\omega_q^3} .
\end{eqnarray}
Away from critical point, we have ${\cal M}_{(-3)}(\hat{O} ) \sim N$, in contrast to ${\cal M}_{(-3)}(\hat{O} ) \sim N^3$ sitting on the critical point. The GFSs ${\cal M}_{(k)}$ with more negative $k$ takes on more singular peaks, shown in Fig.\ref{Mm3-h}. To validate the generalization
to lower order of GFS is significant, we resort to another less investigated model.

The 1D spin-1/2 anisotropic XY
model in a transverse alternating field is defined by
\begin{eqnarray}
H=-\sum_{j=1}^{N}\left\{ \frac{1+\gamma}{2} \sigma_{j}^{x} \sigma_{j+1}^{x}+ \frac{1-\gamma}{2} \sigma_{j}^{y} \sigma_{j+1}^{y}-h_j \sigma_j^z \right\},
\end{eqnarray}
with
\begin{eqnarray}
h_j=h-(-)^j \delta,
\end{eqnarray}
where $\gamma$, $h$ and $\delta$ are anisotropy in the XY plane,
the uniform and alternating magnetic field strengths, respectively. Here periodic boundary conditions are assumed. Using Jordan-Wigner transformation in Eqs. (\ref{JW1}-\ref{JW2}), the model can be diagonalized as:
\begin{eqnarray}
H&=&-\sum_{j}^{N}\{ \gamma (c_{j}^{+}c_{j+1}^{+}-c_{j}  c_{j+1}) +(c_{j}^{+}c_{j+1}-c_{j} c_{j+1}^{+}) \nonumber \\
&-&\left[ h-(-)^j \delta \right](1-2 c_j^+ c_j) \}.
\end{eqnarray}
We introduce the discrete Fourier
transformation of two kinds of fermionic operators in this period-two system:
\begin{eqnarray}
c_{2j-1}=\!\frac{1}{\sqrt{N'}}\sum_{k}e^{-ik j}a_{k},\hskip .2cm
c_{2j}=\!\frac{1}{\sqrt{N'}}\sum_{k}e^{-ik j}b_{k}, \quad
\end{eqnarray}
where $N'=N/2$ and the discrete momenta are given as follows:
\begin{eqnarray}
k=\frac{n\pi}{ N^\prime  }, \quad n= -(N^\prime\!-1), -(N^\prime\!-3),
\ldots, (N^\prime\! -1). \quad
\end{eqnarray}
In terms of vector operators $\Gamma_k^{\dagger} =(a_{k}^{\dagger},a_{-k},b_k^{\dagger},b_{-k})$,
we rewrite it in the BdG form,
\begin{eqnarray}
H &=&\sum_{k}\Gamma_k^{\dagger}{\cal H}_k \Gamma_k,
\end{eqnarray}
where
\begin{eqnarray}
{\cal H}_k =\left(\begin{array}{cccc}
(h+\delta) & 0 & \frac{(1+e^{ik})}{2} & \frac{\gamma(1-e^{ik})}{2} \\
0 & -(h+\delta) & -\frac{\gamma(1-e^{ik})}{2} & -\frac{(1+e^{ik})}{2} \\
\frac{(1+e^{-ik})}{2} &-\frac{\gamma(1-e^{-ik})}{2} & (h-\delta) & 0 \\
\frac{\gamma(1-e^{-ik})}{2} & -\frac{(1+e^{-ik})}{2}  & 0 & -(h-\delta)
\end{array}\right). \label{FT2}
\end{eqnarray}
The diagonalization of the BdG Hamiltonian (\ref{FT2}) has undergone an exhaustive study. More precisely, the diagonalized form is achieved by a four-dimensional Bogoliubov transformation which connects $\{a_k$, $a_{-k}$, $b_k$,
 $b_{-k}\}$ with two kind of quasiparticles,
 \begin{eqnarray}
\left(
\begin{array}{c}
\gamma_{k,1} \\
\gamma_{-k,1}^{\dagger} \\
\gamma_{k,2}\\
\gamma_{-k,2}^{\dagger}
\end{array}
\right)=\hat{U}_{k}^{\dagger} \left(
\begin{array}{c}
a_k  \\
a_{-k}^{\dagger}   \\
b_k  \\
b_{-k}^{\dagger}
\end{array}%
\right), \label{eq:2DXXZ_RDM}
\end{eqnarray}%
where the rows of $\hat{U}_{k}$ are eigenvectors of the
BdG equations,
\begin {eqnarray}
&&[H,\gamma_{k,1}^{\dagger}]=\varepsilon_{k,1}\gamma_{k,1}^{\dagger}, \\
&&[H,\gamma_{k,2}^{\dagger}]=\varepsilon_{k,2}\gamma_{k,2}^{\dagger}.
\end{eqnarray}
Here $\varepsilon_{k,1}$ and $\varepsilon_{k,2}$ are elementary excitations. To this end,
\begin{eqnarray}
H=\sum_{k}\sum_{n=1}^{2} 2 \varepsilon_{k,n} (\gamma_{k,n}^{\dagger}\gamma_{k,n}-\frac{1}{2}),
\end{eqnarray}
where
\begin{eqnarray}
\varepsilon_{k,1}&=&2\left[h^2+\delta^2+\cos^2 \frac{k}{2}+\gamma^2 \sin^2 \frac{k}{2} \right. \nonumber \\ &-&  2 \left.\sqrt{h^2\cos^2 \frac{k}{2} +\delta^2 (h^2 +\gamma^2 \sin^2 \frac{k}{2})} \right]^{1/2}, \\
\varepsilon_{k,2}&=&2\left[h^2+\delta^2+\cos^2 \frac{k}{2}+\gamma^2 \sin^2 \frac{k}{2} \right. \nonumber \\ &+&  2 \left.\sqrt{h^2\cos^2 \frac{k}{2} +\delta^2 (h^2 +\gamma^2 \sin^2 \frac{k}{2})} \right]^{1/2}.
\end{eqnarray}
The ground state energy is thus obtained,
\begin{eqnarray}
E_0 = -\sum_{k} \left(\varepsilon_{k,1} +
\varepsilon_{k,2}\right). \label{E0expression2}
\end{eqnarray}
Since QPTs are caused by nonanalytical behavior of ground-state energy, QCPs correspond to zeros of $\varepsilon_{k,1}$.
Quantum phase boundaries are determined by the equations:
\begin{eqnarray}
 h^2=\delta^2+1 ; \quad   \delta^2=h^2+\gamma^2.
\end{eqnarray}
Apart from a second-order broken-symmetry QPT at $h_{c,1}=\pm \sqrt{\delta^2+1}$ [see Fig.(\ref{D2E0Dh2}a)], it was discovered that a fourth-order QCP occurs at $(h \to 0, \delta= \pm \gamma)$ and $(h = \pm 1, \delta \to 0)$\cite{Deng08} [see Fig.(\ref{D2E0Dh2}b)].
A critical behaviour is characterized by a set of exponents which determine peculiarities
of different ground-state quantities in the vicinity of the critical field. The correlation length $\xi$ $\propto \vert h-h_c \vert^{-\nu}$, and the transverse static susceptibility diverges
as $\chi^z$ $\propto \vert h-h_c \vert^{-\alpha}$, and the energy gap vanishes as $\Delta \propto \vert h-h_c \vert^{\nu z}$, where $\alpha$ and $z$ are the critical exponent of specific heat and the dynamic critical exponent.  The transverse static susceptibility, i.e., the second-order differential of the ground-state energy,  $\chi^z$ = $\partial^2 E_0/\partial  h^2$ diverges at $h_{c,1}=\pm \sqrt{\delta^2+1}$. Figure \ref{M_k_sigmaz_h2} reveals that $\chi^z$ $\propto$ $|h - h_{c,1}|^{-\alpha}$ with $\alpha=0$, that is to say, only a logarithmic divergence in the dependence of $h$ \cite{Oleg05}. The elementary excitation energy occurs at $k=0$ which permits us to get for the energy gap in the vicinity of the critical points, $\Delta$  $\propto$ $|h - h_{c,1}|$, implying $\nu z$ = 1.
When $\gamma$ $>$ 0, generic QCPs belong to the two-dimensional Ising universality class with critical exponents $\nu$ = 1, $z$ =1.
Nevertheless,
we find that around $h_{c,2}=0$, $\chi^z$ $\propto$ $|h - h_{c,2}|^{-\alpha}$ with $\alpha=-2$, i.e., $\chi^z$ does
not diverge at QCP at $h_{c,2}$ and $\partial^2 \chi^z/\partial  h^2$ exhibits a logarithmic divergence at $h_{c,2}$, implying a fourth-order QPT. The energy gap occurring at $k=\pi$ in the vicinity of the critical points behaves as $\Delta$  $\propto$ $|h|^2$, implying $\nu z$ = 2. The critical exponents $\nu=2$, $z=1$ are identified via careful analysis, corresponding to an alternating universality class. The obtained critical exponents confirm one of the scaling relations in QPT: $2-\alpha=(d+z)\nu$.
\begin{figure}[h]
\includegraphics[width=8cm]{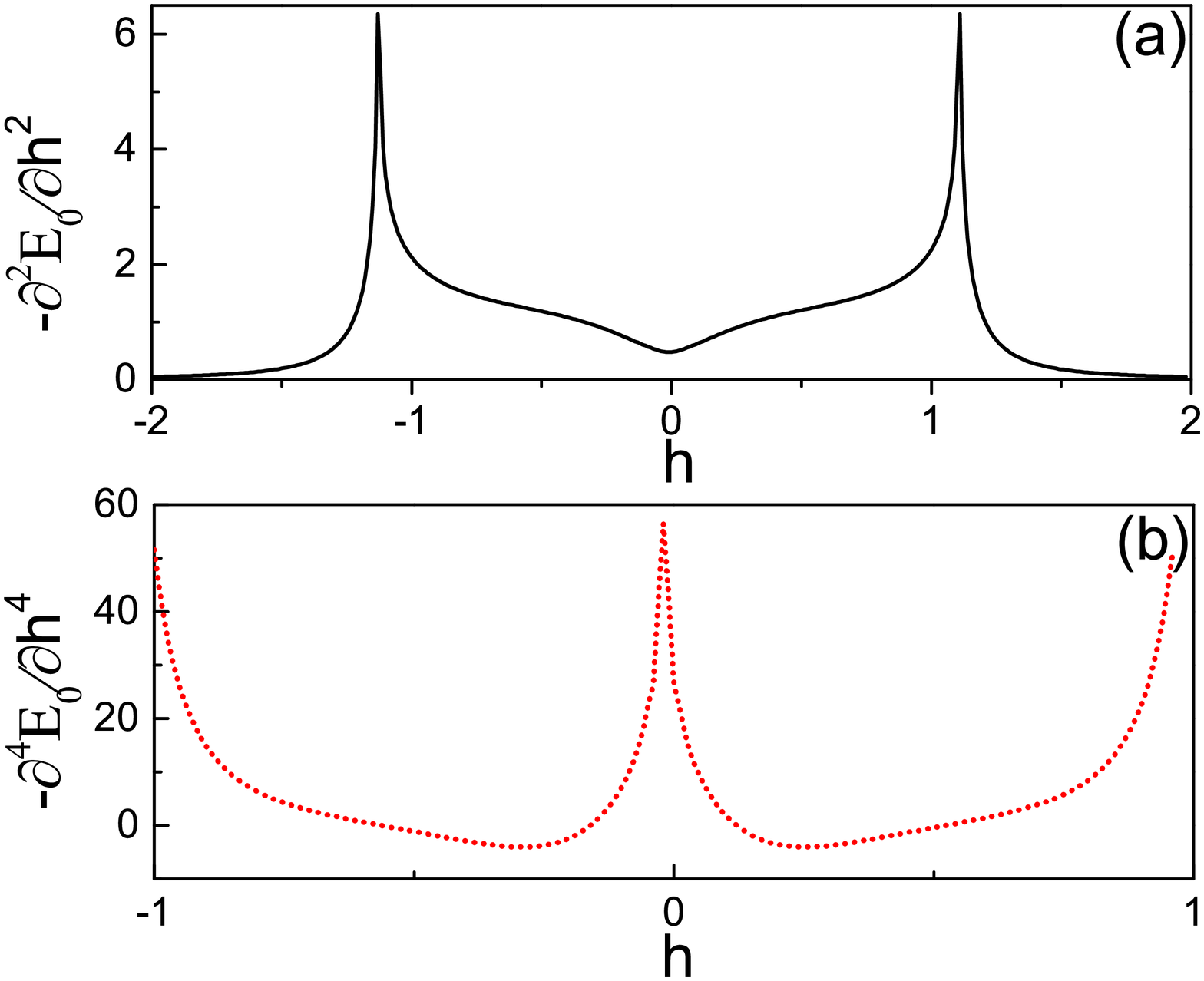}
 \caption{(Color online) Derivatives of ground-state energy with respect to $h$ in the 1D spin-1/2 anisotropic XY model under a transverse alternating field. (a) -$\partial^2 E_0/\partial h^2$; (b) -$\partial^4 E_0/\partial h^4$. Parameters are as follows: $\gamma=0.5$, $\delta=0.5$, $N=8000$.} \label{D2E0Dh2}
\end{figure}

By Eq. (\ref{Mkdefinition}), the GFSs of different orders are showcased in Fig. \ref{M_k_sigmaz_h2}. We find that the critical points $h_{c,1} $ have been already captured by the singular peak of ${\cal M}_{(-1)}(\hat{\sigma}^z)$, while only ${\cal M}_{(-3)}(\hat{\sigma}^z)$  can detect the anomaly around $h_{c,2}$ with a singular peak. We can observe that a similar behavior takes place for the GFSs of $\hat{O}_{\rm MF}$; see Fig. \ref{M_k_OMF_h}. We can speculate more pronounced peaks of  ${\cal M}_{(k)}(\hat{O})$ will emerge for $k<-3$.

\begin{figure}[h]
\includegraphics[width=8cm]{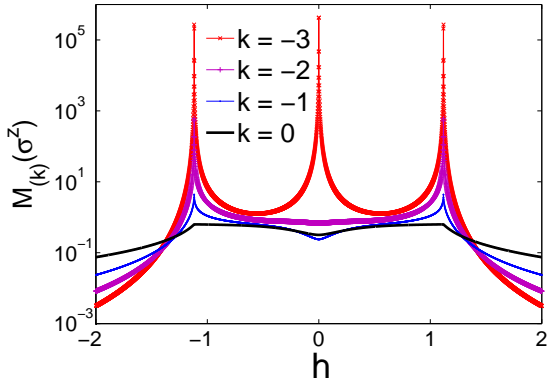}
 \caption{(Color online) The GFSs of $\hat{\sigma}^z$ with different orders in the 1D spin-1/2 anisotropic XY
model under a transverse alternating field. Parameters are as follows: $\gamma=0.5$, $\delta=0.5$.} \label{M_k_sigmaz_h2}
\end{figure}
\begin{figure}[h]
\includegraphics[width=8cm]{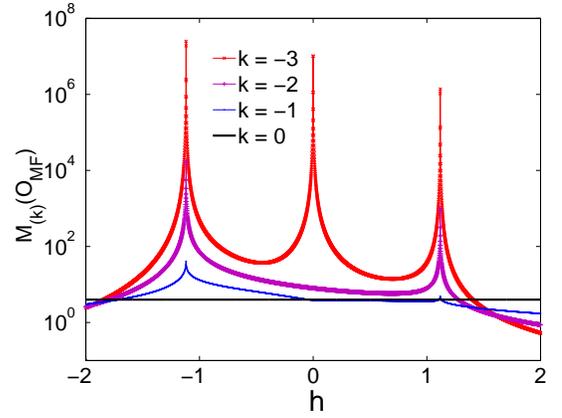}
 \caption{(Color online) The GFSs of $\hat{O}_{\rm MF}$ with different orders in the 1D spin-1/2 anisotropic XY
model under a transverse alternating field. Parameters are as follows: $\gamma=0.5$, $\delta=0.5$.} \label{M_k_OMF_h}
\end{figure}

\section{Conclusions and discussion}
\label{sec:conclusion}
In this paper, we proved the equivalence of two forms of the QFS between Eq.(\ref{chiperturbationform}) and Eq.(\ref{chipartialform}). We show in a sense that the QFS is neither a pure descendant of the concept of fidelity, nor a response of the perturbed Hamiltonian. Instead, it connects the critical properties with many relevant operators. We discuss the QFS of $z$-component magnetization $\hat{\sigma}^z$ and momentum space Majorana operator $\hat{O}_{\rm MF}$ in the 1D transverse Ising model, which presents similar critical behaviors. The QFS can be generalized to moments ${\cal M}_{(k)}$ of the dynamic form factor, which are experimentally accessible. In the vicinity of a critical point, a lower-order GFS ${\cal M}_{(k-1)}$ ($k \le -2$) will exhibit stronger singular behavior than that of ${\cal M}_{(k)}$. Considering that a singularity in the second derivative of energy seizes a second-order QPT, it is supposed that ${\cal M}_{(-k+1)}$ of relevant operators is requisite to detect the criticality for a $k$-order phase transition. We can observe that ${\cal M}_{(-1)}$, namely, magnetic susceptibility $\chi$, has detected the second-order QPT in 1D transverse Ising model, and lower-order GFSs display much stronger peaks. Nevertheless, for the fourth-order QPT occurring in the 1D spin-1/2 anisotropic XY model under a transverse alternating field, we find that ${\cal M}_{(-2)}$ does not show any anomaly in the vicinity of QCP, while ${\cal M}_{(-3)}$ or lower-order GFSs can capture the singularity. This finding infers that the QFS may not predicate the the existence of the Berezinskii-Kosterlitz-Thouless QCP, which affirms our argument in
Ref.[\onlinecite{You07}] from another perspective.

In the formulas (\ref{Mkdefinition}) and (\ref{Mktemperature3}), the structure change is encoded in the transition matrix elements $\vert \langle \Psi_l \vert \hat{O} \vert \Psi_0 \rangle \vert^2$ of an observable $\hat{O}$ between the ground and excited states.
The scope of the GFS can be reaily applied to the related solvable models like the transverse XY chain \cite{Zanardi06}, dimerized XX model \cite{Venuti10}, the BCS Hamiltonian \cite{Anderson58}, the Harper chain \cite{Satija}, the Kitaev model \cite{Kitaev06} and compass chain \cite{You12,You14a,You14b}. Also, the concept of moment is easily extended to finite temperature \cite{Sirker10}.
Besides in those solvable models or very small systems,
the matrix elements can also be potentially updated and evaluated via the prevailing numerical techniques, like the Monte Carlo methods and transfer-matrix renormalization group technique. More importantly, in experiment the frequency resolved structure factor is easily measured via the well known techniques, such as neutron scattering, angle resolved photoemission spectroscopy (ARPES) techniques, etc. For instance, the 1D quantum transverse Ising model was realized in the
insulating ferromagnet CoNb$_2$O$_6$ under strong transverse magnetic fields and the spin dynamics
was measured by neutron scattering \cite{Coldea10}. By Eq. (\ref{Mkdefinition}), the GFS can be derived from the knowledge of the dynamic structure factor.

\acknowledgments
W.-L.Y. thanks Shi-Jian Gu for the hospitality and insightful discussion during the visit in The Chinese University of Hong Kong. W.-L.Y. also thanks Zeng-Qiang Yu and Guang-Hua Liu for their stimulating discussion.
This work is supported by the National Science Foundation of China
(NSFC) under Grant No. 11474211 and the Natural Science Foundation
of Jiangsu Province of China under Grant No. BK20141190.

\end{document}